\documentclass{optica-article}

\journal{opticajournal} 

\articletype{Research Article}

\usepackage{indentfirst}
\usepackage{xr}
\usepackage{lineno}
\usepackage{bm}
\usepackage{soul}
\usepackage{cite}
\begin{document}

\title{High-Performance, Efficient, Low-Fresnel Number Focusing Metamirror in the D-Band}

\author{Fahim Ferdous Hossain,\authormark{1}  Yun-seok Choi,\authormark{2} Abul K. Azad\authormark{3} AND John F. O'Hara\authormark{4,*}}

\address{\authormark{1,4}School of Electrical and Computer Engineering, Oklahoma State University, Stillwater, OK, USA\\
\authormark{2,3}Center for Integrated Nanotechnologies, Los Alamos National Laboratory, Los Alamos, NM, USA.}

\email{\authormark{*}oharaj@okstate.edu} 


\begin{abstract*} 
Metasurface-based optical components are becoming increasingly important due to their unparalleled ability to shape and manipulate electromagnetic waves across a wide range of frequencies, from microwave and terahertz to visible light. In particular, planar meta-mirrors with high efficiency and broad operational bandwidth are expected to play a significant role in shaping channel characteristics for 6G communication links due to their reduced SWaP (size, weight, and power) and integrability. However, achieving large fractional bandwidth and high efficiency simultaneously using metasurface-based mirrors or reflectors have been challenging. In this work, highly efficient broadband focusing meta-mirror has been designed, fabricated and experimentally demonstrated. The device operates in the D-Band (110-170 GHz) and beyond (up to 200 GHz), and exhibits unmatched performance in terms of combined fractional bandwidth (58\%) and energy efficiency, exceeding 53\% over the entire bandwidth with a peak of 77.4\% at frequency 135 GHz. The high performance and advantageous design tradeoffs of this meta-mirror are discussed in terms of the principles of low Fresnel number optics.

\end{abstract*}

\section{Introduction}
Wireless data traffic has been rapidly growing over the last few decades due to the emergence of applications such as virtual reality (VR), artificial intelligence (AI), three dimensional media, edge computing, and the Internet of Everything (IoE), yet current wireless networks are unable to match the swiftly increasing technical requirements \cite{Jang_2020, Shaoqian_2019}. The current 5th generation (5G) wireless systems will be supplanted by sixth generation (6G) systems to address these requirements, including higher data rate (terabits per second - Tbps) and lower latency (sub-ms) \cite{Jang_2020, Shaoqian_2019, Nie_2020, Tufvesson_2021}. In support of high data rate, terahertz carrier frequencies are a critical enabler of the 6G physical layer, due to the availability of much larger bandwidths \cite{Jang_2020, Tong_2021}. Here, we adopt the common definition of the terahertz band, spanning 0.1 to 10~THz \cite{Tong_2021, Kurner_2023, AlNaffouri_2022}. 

However, the use of terahertz waves also brings numerous challenges, particularly high free-space path loss (FSPL) and the consequent requirements for narrower beam forming and more capable beam steering. Several technologies have been envisioned to meet this need, and one that has drawn considerable interest is the intelligent reflecting surface (IRS) \cite{Jang_2020,Tufvesson_2021,Leung_2024}. Intelligent reflecting surfaces can be realized using antennas spaced on the order of half a wavelength or by metamaterials---specifically metasurfaces~\cite{Shamai_2020}. Metasurfaces are planar, two-dimensional metamaterials~\cite{Nanfang_Yu_2016}, and are made of sub-wavelength unit cells (meta-atoms) usually arranged periodically. Their functionality arises from the phase and amplitude shifts imparted to electromagnetic waves by these meta-atoms. Since they are generally straightforward to fabricate, compact, and low in loss~\cite{Nanfang_Yu_2016, Ping_Tsai_2018}, metasurfaces are a good foundation for IRSs. Various types of devices based on metasurfaces have been demonstrated, including invisibility cloaks~\cite{Erping_2020, Ndao_2019}, absorbers~\cite{Shen_2018, Padilla_2017, Lonkar_2014}, vortex beam generators \cite{Gaburro_2011}, holographic devices ~\cite{Xiangang_Luo_2019, Kildishv_2013,Weili_2016, Chunmei_2017}, focusing \cite{Ren_Jie_Lin_2019, Shuming_Wang_2018, Din_ping_Tsai_2017, Zentgraf_2012,Balakin_2021, Baihong_2020},special beam generation~\cite{Xinke_2013,Zhenwei_2016, Huan_2017, Sen_2017, Minggui_2018, zHongzhen_2018, Raghu_2019}, polarization control~\cite{Guohua_2020, Yunfei_2020}, and more. Wave manipulation in the terahertz regime is particularly interesting as 6G research converges into the D-band frequency regime and beyond. 

 Metasurface approaches to focusing are important since they underpin next-generation planar ultra-compact beam forming and beam shaping or collimating devices. There have been a large number of numerical demonstrations of artificial focusing devices but far fewer experimental demonstrations. A polarization conversion-dependent metalens and a lens array with split ring resonators as structural units operating in 0.5--0.9~THz was experimentally demonstrated in \cite{Xueqian_2015}. Dielectric cross resonators were used to fabricate a lens operating at 3.11~THz having 24\% efficiency \cite{Ma_2017}. Dielectric cube shaped resonators have been utilized to demonstrate a polarization dependent lens operating at 2.53~THz \cite{Gaofeng_2018}. Employing polarization conversion, a bifocal cylindrical metalens operating at 0.6--0.8~THz with an energy efficiency higher than 32\% and a cylindrical metalens operating within 0.308--0.381~THz with efficiency above 52\% were demonstrated in \cite{Chunrui_2024}. A vanadium dioxide based switchable focusing metalens opearating within 0.5--0.68~THz having an efficiency $<30$\% was exhibited in \cite{Shixiong_2022}. Finally, a polarization dependent multiple foci metalens operating at 0.7~THz was demonstrated to have an efficiency of 27.6\% \cite{Binbin_2024}.

There have been several demonstrations of focusing metasurface \emph{reflectors} in different terahertz bands as well. A focusing reflectarray operating with a center frequency of 1~THz, having a 3~dB fractional bandwidth of 23.3\% for transverse electric (TE) polarized excitation and 23.9\% for transverse magnetic(TM) polarized excitation was experimentally realized in \cite{Withayatchumnankul_2019}. The efficiency at the center frequency was 71.9\% and 71.0\% for TE and TM excitation respectively \cite{Withayatchumnankul_2019}. A focusing folded meta device having a 3~dB gain bandwidth of about 7\% and operating in the 1.024 to 1.1~THz was manufactured and characterized in \cite{Stella_2022}. A dielectric resonator based reflectarray having a 3~dB bandwidth of 18\% and a center frequency of 1~THz was experimentally validated in \cite{Headland_2016}. Focusing metasurfaces having a maximum efficiency of 80\% at the center frequency of 0.35~THz exhibited 3~dB bandwidth of 19\% and 15\% for 1--spot focusing and 4--spot focusing (beamsplitting) in \cite{Astafev_2015}. Active reflecting focusing metasurfaces have found practical application in quantum cascade lasers in the 3.2-3.5~THz region \cite{Dagun_2016}. A frequency controlled focus scanning metasurface operating in 0.225-0.3~THz was experimentally demonstrated in \cite{Bai_2017}. A metal-insulator-metal structure based focusing metasurface operating in 200-300~GHz range with a fractional bandwidth of 40\% was numerically investigated \cite{Renshuai_2019}.  Previous experimental research works on focusing terahertz metasurfaces have been summarized in Table~S1 in Supplement 1.

It is apparent from the literature that it is very difficult to achieve both high fractional bandwidth and efficiency at the same time in metasurface based devices. However, it is also clear that such devices will become increasingly important. In this paper, we present a metal-insulator-metal (MIM) based metamirror operating in the D-band, between 0.11--0.2~THz with an efficiency above 53.5\% over the entire bandwidth, peaking at 77.4\%. We experimentally characterize its focusing behavior and compare it to an ideal focusing mirror, showing close agreement between them. To the best of our knowledge, the aforementioned performance parameters have not been experimentally demonstrated so far. Furthermore, we offer some discussion on how this performance was made possible by leveraging the design benefits of low Fresnel numbers optics.

\section{Design of the metamirror}

The focusing metamirror presented in this paper was designed as a flat, or planar, metasurface. To implement focusing behavior the surface must impart spatially-dependent phase shifts ranging from $\phi = [ 0 :2\pi] $~radians over its entire area. Functionally, the metamirror should transform incident plane waves into converging spherical waves centered at the focal length $f$. We can imagine a three-dimensional coordinate system (shown in Figure~S1 in Supplement 1) where the metamirror is in the $XY$ plane centered on the origin $(0,0,0)$. Further, the ideal phase shift imparted by the metamirror at any point $(x,y,0)$ is
\begin{equation}
\Delta\phi(x,y) = \frac{2\pi}{\lambda}\left(\sqrt{x^2+y^2+f^2}-f\right) ,
\label{eqn:1}
\end{equation}
relative to the phase at the origin, which is assumed to be zero. 

\begin{figure}[h!]
  \centering
  \includegraphics[width= 7.5 cm]{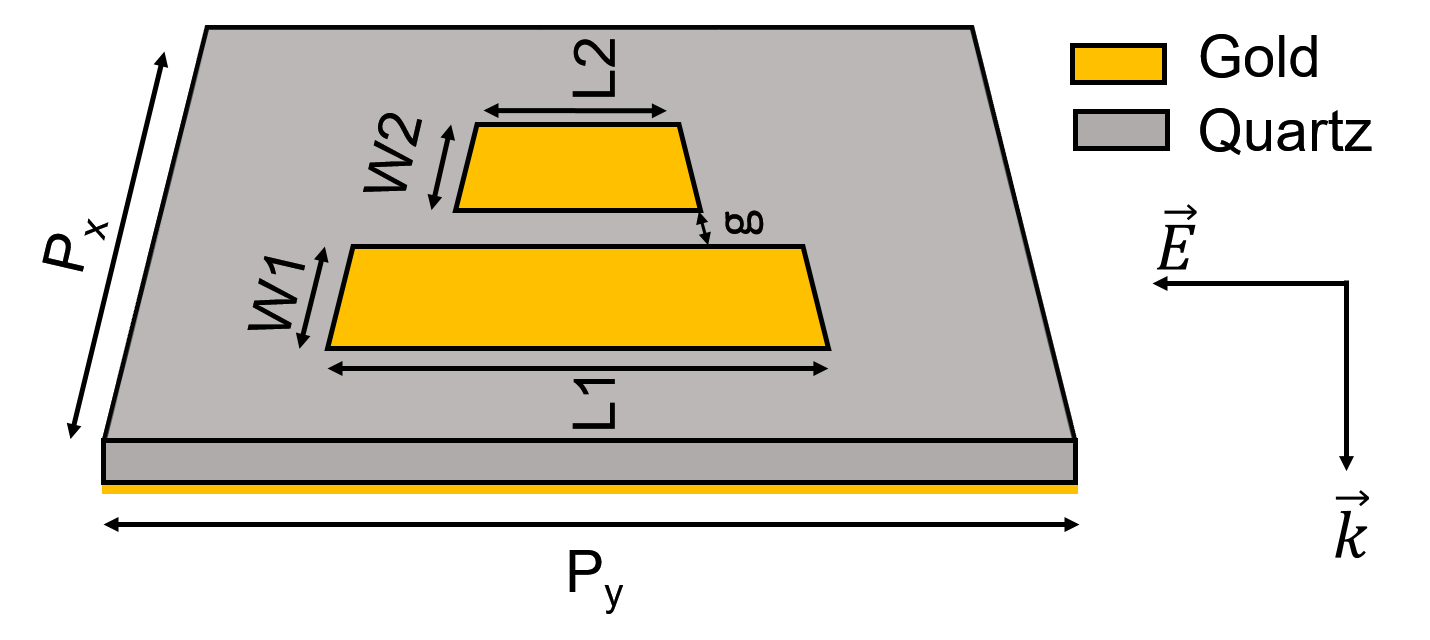}
  \caption{Schematic of a sample unit cell. $L1$, $W1$, $L2$, $W2$, and $g$ were tuned to vary the reflection $S_{11}$. $P_{x}$ and $P_{y}$ are the unit cell dimensions. $\overrightarrow{E}$ and $\overrightarrow{k}$ denote the directions of the electric field and wave vector used in simulation of the unit cells.}
  \label{fig:unit_cell}
\end{figure}

Having the required phase shifts, the design challenge turns to implementation. To achieve high reflection efficiency, metasurfaces are often implemented in a MIM configuration, where the metasurface and a metal backplane are fabricated on either side of an insulating dielectric slab. The varying phase shifts can be achieved with a combination of different sub-wavelength, metallic, resonant structures. The metamirror in this work was designed and simulated using high conductivity metal and a quartz insulator. The non-magnetic quartz had a relative permittivity of $\epsilon_r=4.435$~\cite{Gregory_terahertz_2021}, a loss tangent of $\tan\delta=37\times10^{-6}$~\cite{Gregory_terahertz_2021}, and zero conductivity. The metal conductivity was set to be $\sigma = 3.77\times10^{7}$~S/m~\cite{ordal_1983}. The unit cells consisted of metallic patches in pairs, whose lengths, widths, and separations have been varied to achieve desirable responses in reflection magnitude and phase. Additional design considerations accounted for the Fabry-Perot cavity formed within the insulator. The resulting overall reflected phase and magnitude is finally expressed as the complex S-parameter, $S_{11}$. A commercial finite element software (COMSOL) was used to obtain the different $S_{11}$ values of a variety of different unit cells excited by the $y$-polarized (parallel to $P_{y}$ in Figure~\ref{fig:unit_cell}) normally incident plane wave. A sample unit cell is depicted in Figure~\ref{fig:unit_cell}. The length of the square unit cells was fixed at 650~$\mu$m over the entire metamirror area. 
Having a large set of parameterized simulations, a metamirror design was found using an optimization method that sought to maximize $|S_{11}|$ and optimally reproduce the phase according to Eq.~\ref{eqn:1} over the operational bandwidth. A 200~$\mu$m thick quartz insulator was found to be favorable for practical implementation and metamirror performance. From a pool of simulations, a set of 9 optimal unit cells was chosen to populate nine different annular regions of a 90~mm diameter circular surface to realize a focusing metamirror intended to behave like a conventional, concave, mirror with a geometric focal length of $f=500$~mm. The width of each annular region was 5~mm. Each of the chosen unit cells exhibited $|S_{11}|>0.9$ in the 110~GHz to 200~GHz bandwidth. The dimensions, reflection magnitude and phase responses  of the rectangular metal patches of the chosen unit cells have been included in Supplement 1.

\begin{figure}[h!]
  \centering
  \includegraphics[width= 8 cm]{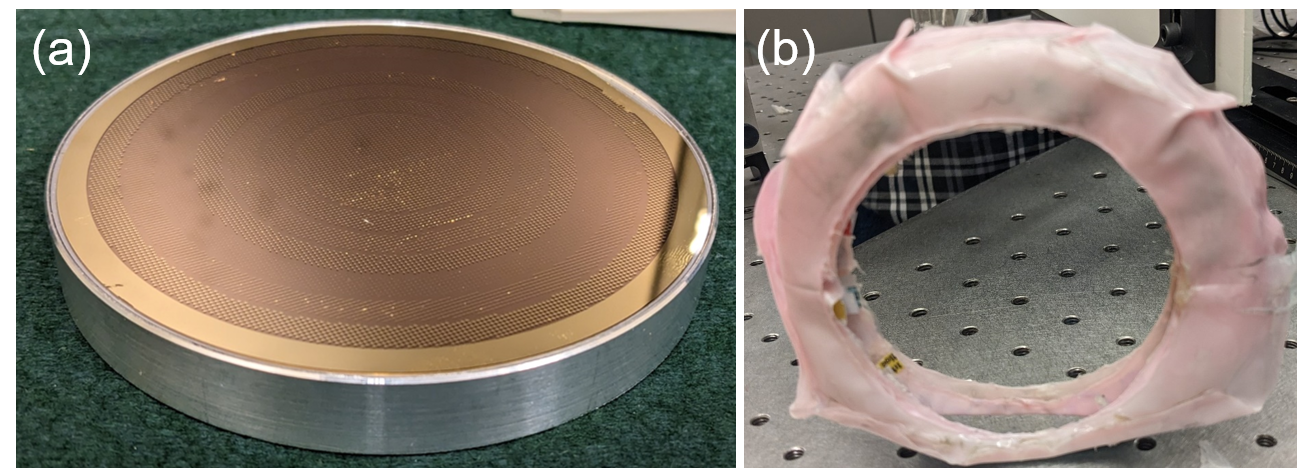}
  \caption{(a) Fabricated metamirror, (b) ideal focusing mirror: the outer pink plastic covering encases terahertz absorbing material to reduce the effective aperture radius from 50~mm to 45~mm}.
  \label{fig:fabricated_sample}
\end{figure}

With a suitable design and promising simulated performance, the metamirror was realized using conventional microfabrication techniques. Heidelberg MLA150 was employed to pattern meta-atoms on the photoresist (AZ 5214) coated, 200-um thick quartz substrate. The patterns were developed by MIF319. A 5~nm Ti layer and a 200~nm Au layer were deposited by electron beam physical vapor deposition on the patterned front side and unpatterned backside of the substrate. The sample was soaked in Remover PG overnight for the liftoff process. The fabricated metamirror is shown in Figure~\ref{fig:fabricated_sample}(a).  

\section{Experimental Setup}
\begin{figure}[h!]
  \centering
  \includegraphics[width= 7.5 cm]{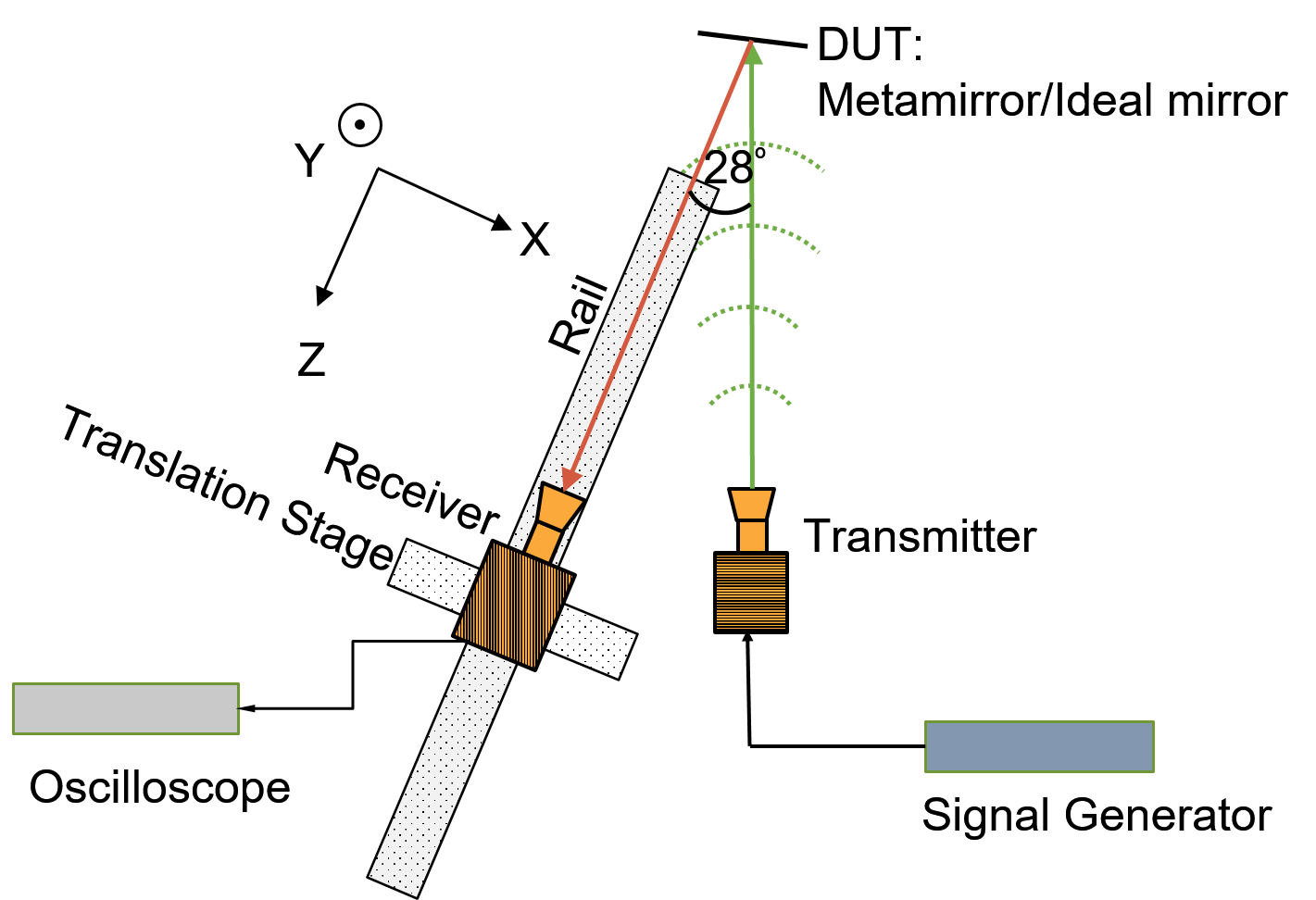}
  \caption{Schematic of the experimental setup. The green arrow line is the incident beam axis and the red arrow line is the reflected beam axis.}
  \label{fig:One_to_one}
\end{figure}

To quantify whether the realized metamirror behaved as intended, a conventional, concave mirror (or ``ideal'' focusing mirror)(Figure~\ref{fig:fabricated_sample}(b).) having a diameter of 90~mm (same as the metamirror) and a geometric focal length of 500~mm was acquired and also measured in the same imaging setup. The focusing performance of the mirrors was experimentally measured in a system illustrated in Fig~\ref{fig:One_to_one}. In this nearly one-to-one imaging configuration, the distance between both the terahertz transmitter or receiver and the device under test (DUT, i.e. metamirror or ideal focusing mirror) were similar. To be precise, the transmitter-to-DUT distance was $s_o=895$~mm. The receiver then explored the region near the image plane, which should be located at $s_i = (1/f-1/s_o)^{-1}=1133$~mm according to the geometric optics approximation. The mirrors' surface normal vectors at $(x=0,y=0)$ were also oriented at an obliquity angle of $14^{\circ}$ relative to the incident and reflected beam axes. This arrangement enables a compact arrangement of the equipment and produces nearly spherical phase fronts at the DUT.

A continuous-wave, tunable transmitter illuminated the DUT via a horn antenna with a divergence angle of approximately $13^{\circ}$. The s-polarized (TE), approximately Gaussian, transmitted beam  fully illuminated the DUT with about 3~dB maximum power variation over the DUT area. Absorbers were placed around the DUT perimeters to eliminate stray reflections. Henceforth, we refer to the coordinate system in Figure ~\ref{fig:One_to_one}, where the +Z axis is coincident with the reflected beam axis. The receiver utilized a conical horn setup equivalent to the transmitter and was mounted onto two motorized linear translation stages, which permitted automated sampling of the image plane in the vertical (Y-axis) and horizontal (X-axis) directions. This entire receiver assembly was then affixed to a carrier/rail mechanism to enable manual selection of the Z-position of the measured image plane. In this way, the receiver could sample the entire image volume. 

Single-frequency tones of 111, 117, 125, 135, 145, 155, 165, 175, 185, 195 and 200~GHz were selected to measure the DUTs. The waves reflected from the DUTs were collected by the receiver, then sampled and stored using a digital storage oscilloscope (DSO). The DSO provided the signal strength in dBm, which was recorded for every frequency of interest. 

This measurement process was repeated for every point in the various XY, XZ and YZ planes. The total image volume explored was $x=\pm23$~mm, $y=\pm23$~mm, and $z = [331,929]$~mm. However for frequencies $\ge170$~GHz - the lower Z-limit was $z=429$~mm, since there were few relevant data features below this distance. Each XY plane consisted of a $19\times19$ square grid of measured points. Images were measured in multiple XY planes with adjacent planes having a spacing of 50~mm. Data were also recorded at points within the XZ and YZ planes. The XZ and YZ planes covered the same spatial ranges listed above using 25 sample points along the Z axis direction and 19 points along the X or Y directions.
\begin{figure}[h!]
  \centering
  \includegraphics[width=12.5 cm]{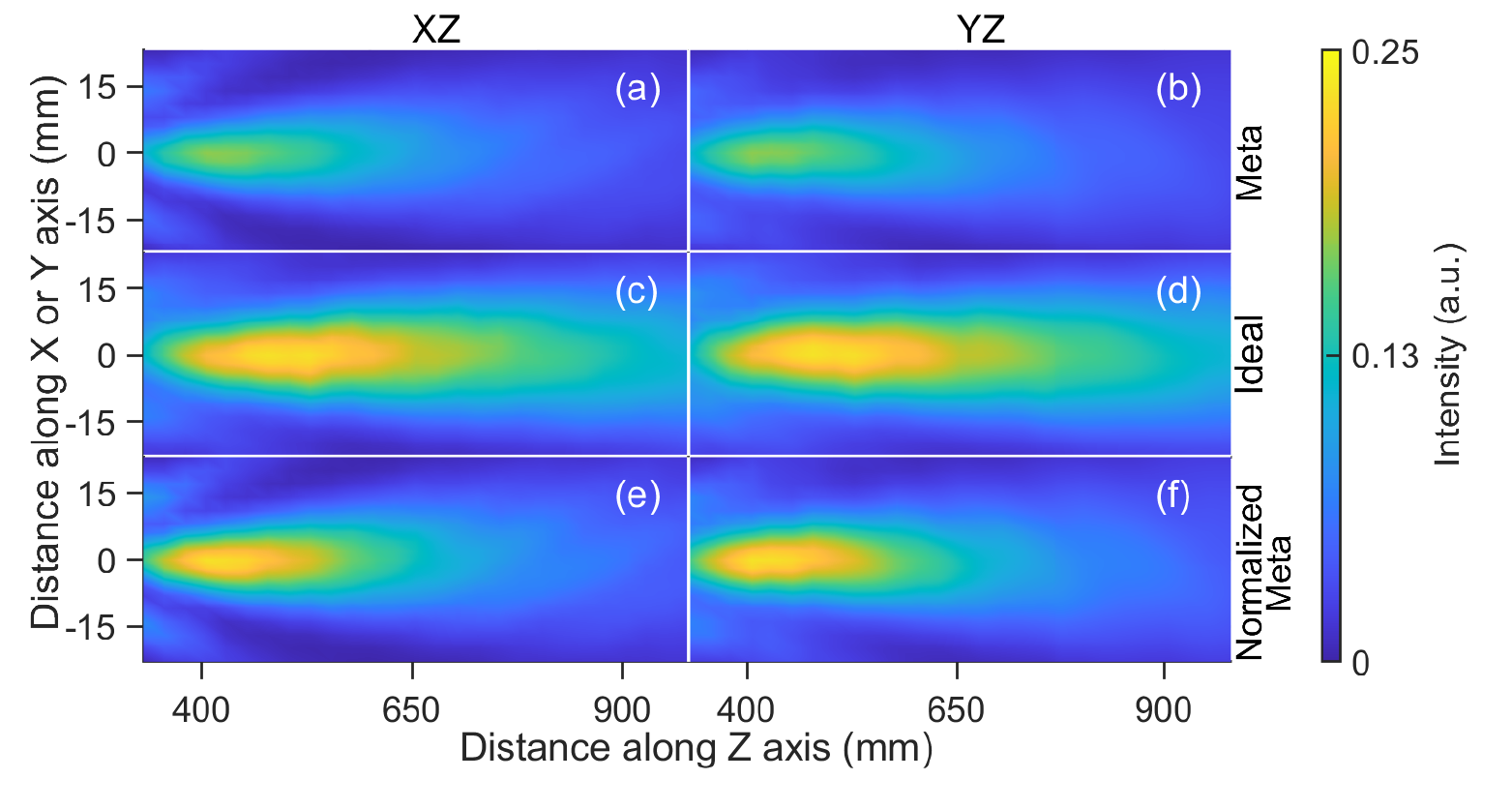}
  \caption{Intensity profile at 111~GHz: XZ and YZ planes of metamirror (a,b) and ideal mirror (c,d). Metamirror profiles are repeated in (e,f) except normalized to the peaks of ideal profiles.}
  \label{fig:XZ_YZ_1}
\end{figure}
\begin{figure}[h!]
  \centering
  \includegraphics[width= 12.5 cm]{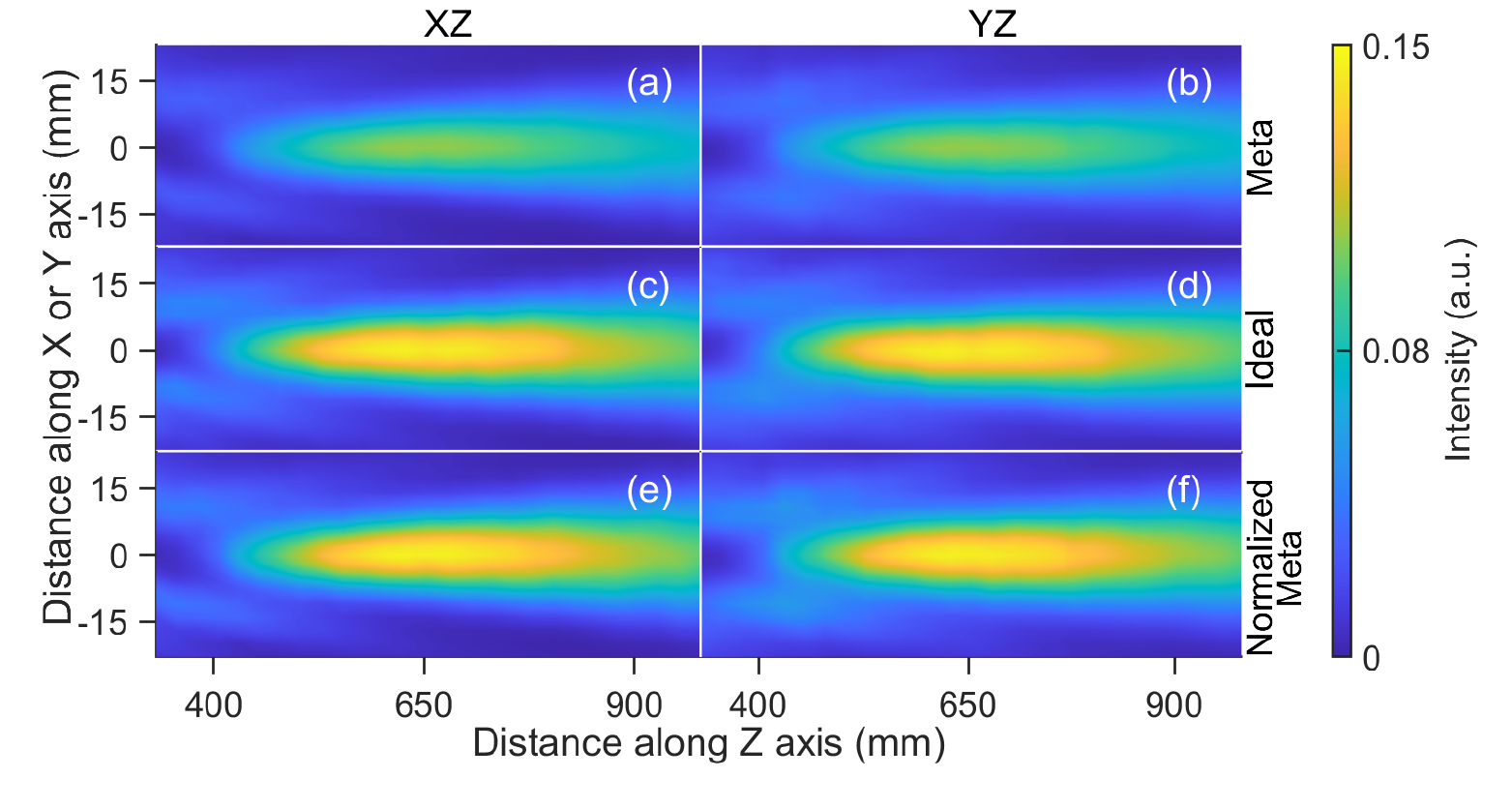}
  \caption{Intensity profile at 165~GHz: XZ and YZ planes of metamirror (a,b) and ideal mirror (c,d). Metamirror profiles are repeated in (e,f) except normalized to the peaks of ideal profiles.}
  \label{fig:XZ_YZ_2}
\end{figure}
\begin{figure}[h!]
  \centering
  \includegraphics[width= 12.5 cm]{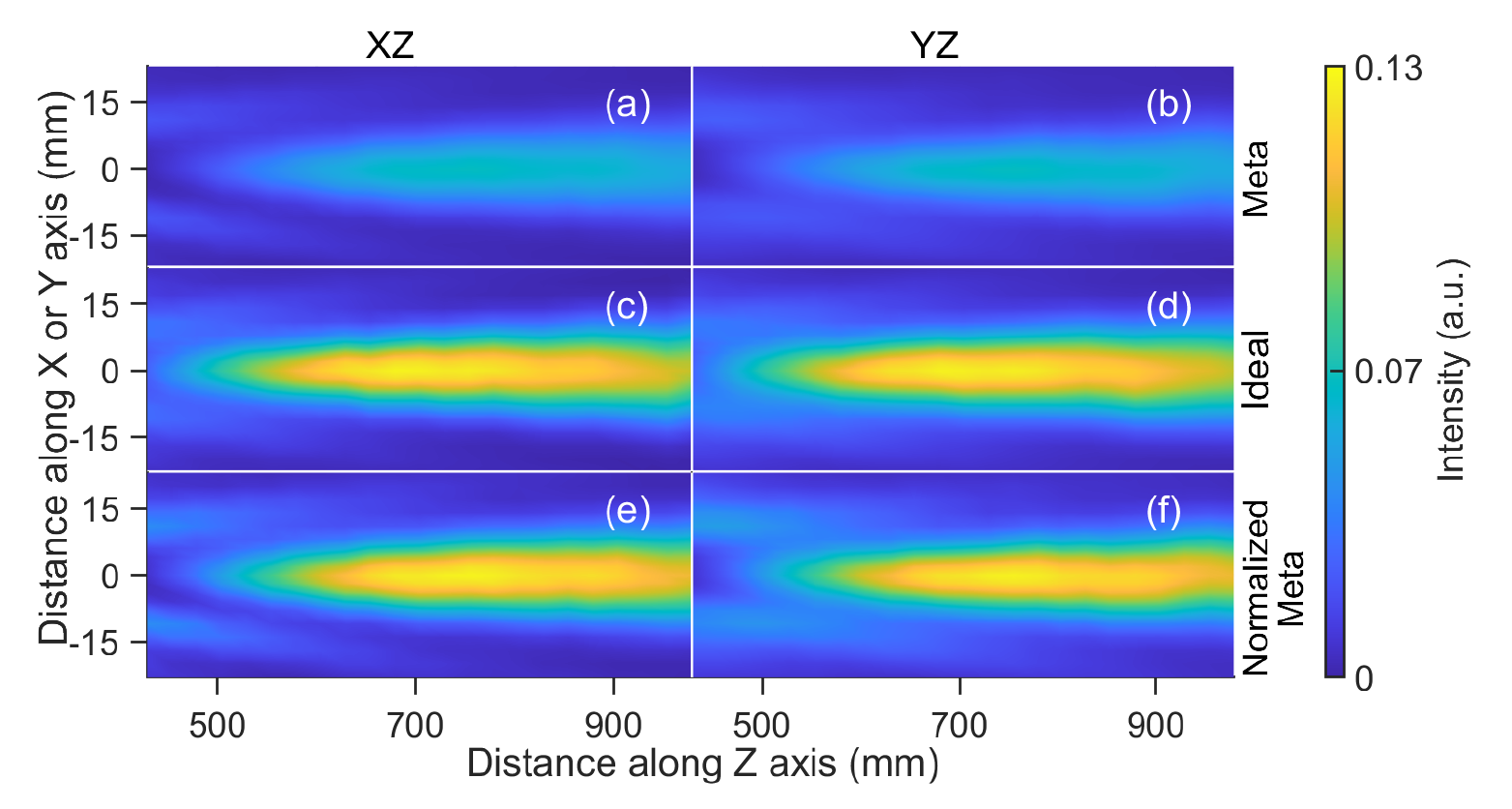}
  \caption{Intensity profile at 200~GHz: XZ and YZ planes of metamirror (a,b) and ideal mirror (c,d). Metamirror profiles are repeated in (e,f) except normalized to the peaks of ideal profiles.}
  \label{fig:XZ_YZ_3}
\end{figure}

\begin{figure}[h!]
  \centering
  \includegraphics[width= 12.5 cm]{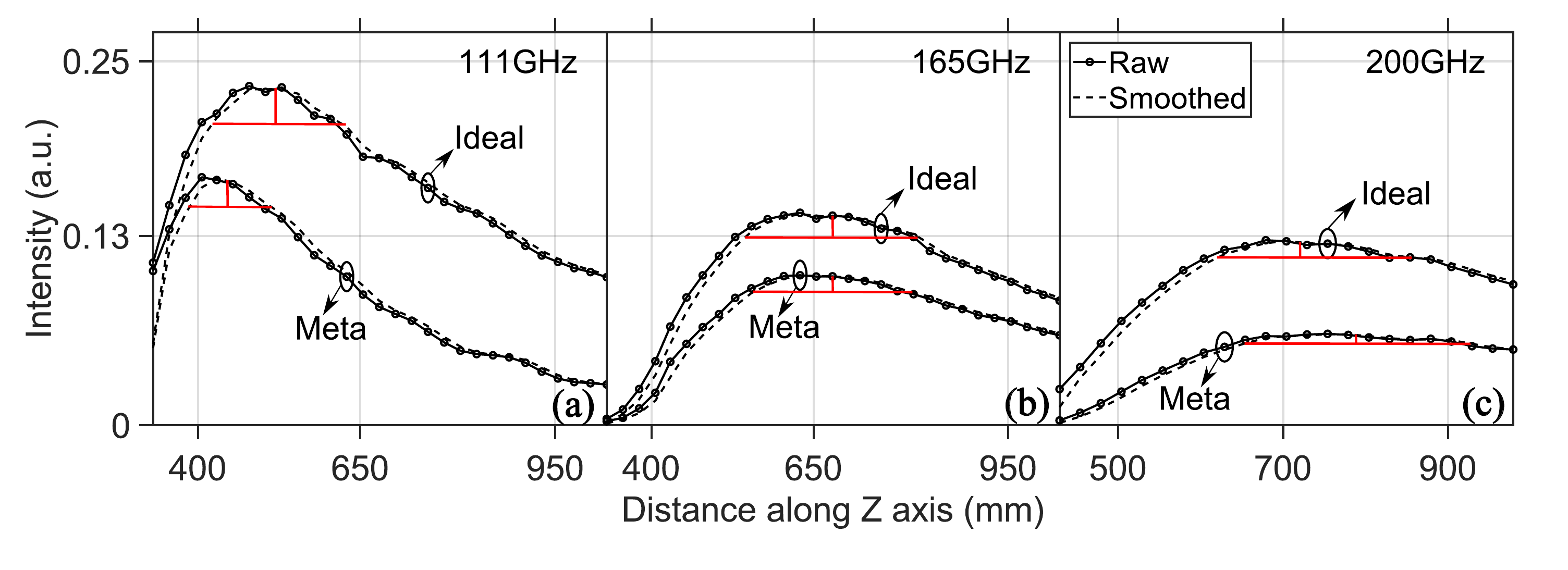}
  \caption{Intensity profiles for both meta and ideal focusing mirrors along Z axis for a) 111~GHz, (b) 165~GHz, and (c) 200~GHz. The horizontal red line segments connect the intensity points which are 90\% of the peak intensity. The vertical line segments denote the position of the image distance. }
  \label{fig:Peak_Value_XZ}
\end{figure}

\begin{figure}[t]
  \centering
  \includegraphics[width= 8.5 cm]{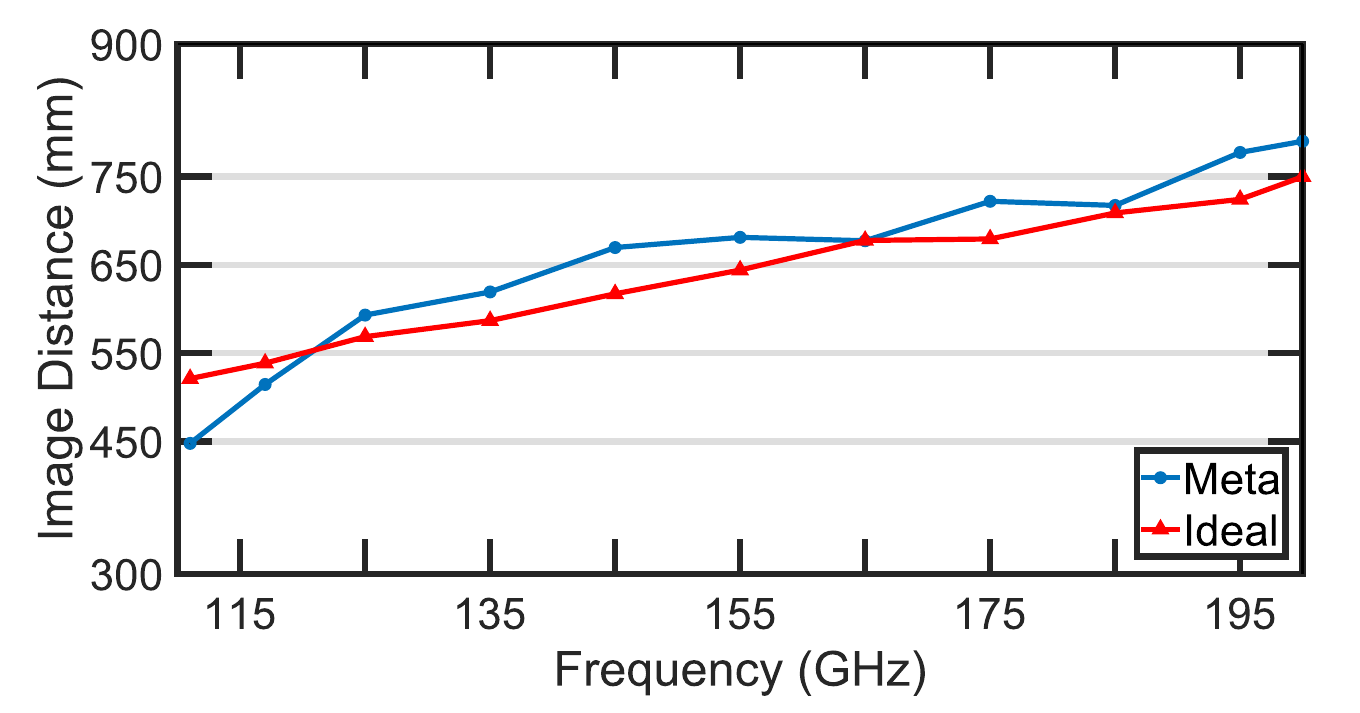}
  \caption{Image plane distance depending on frequency}
  \label{fig:Image_Distance}
\end{figure}

\section{Results}
The measured data reveals the metamirror performance, by which we mean the match between the metamirror and ideal focusing mirror in terms of image distance, beam waist at image distance, and efficiency over the entire operating bandwidth. Figs.~\ref{fig:XZ_YZ_1}, \ref{fig:XZ_YZ_2}, and ~\ref{fig:XZ_YZ_3} show the intensity profiles at the XZ and YZ planes for both the metamirror and the ideal focusing mirror for frequencies 111~GHz, 165~GHz and 200~GHz respectively. For ease of comparison, the intensity profiles of the metamirror have also been normalized to the peak values of the respective ideal focusing mirror intensity profiles and are presented in subplots (e) and (f). For a particular frequency, the intensity profiles in the XZ and YZ planes are very similar in structure indicating that the metamirror design produces the same general focusing behavior as the ideal focusing mirror, particularly in the middle of the operating band. The tubular shape of the image, which narrows and elongates as frequency increases, is commensurate with the expectations of low Fresnel number focusing optics.

Closer examination reveals differences between the ideal focusing mirror and metamirror. First, the XZ and YZ image profiles of the metamirror are weaker in intensity than those of the ideal focusing mirror, suggesting a diminished power efficiency. This is not surprising since the ideal focusing mirror would be expected to be almost perfectly efficient. Additionally, we observe that the first intensity nulls surrounding the main image tube are more pronounced in the metamirror profiles at 111~GHz and 200~GHz. This is likely due to non-idealities in the $S_{11}$ parameters of the metasurface, which were greatest at the band edges. Third, from Figs.~\ref{fig:XZ_YZ_1} - \ref{fig:XZ_YZ_3} we see that the Z-position of the peak intensity of the metamirror image shifts relative to that of the ideal focusing mirror. At 111~GHz, the metamirror image is closer to the mirror. Then, the images of both mirrors appear at the same position at 165~GHz. Finally, at 200~GHz the metamirror image is farther from the mirror. To quantify this evolution, Figure~\ref{fig:Peak_Value_XZ}, plots the intensity values on the Z axis for three different frequencies for both mirrors -- smoothed versions of these curves generated by a moving average filter are also plotted. The tubular image makes it ambiguous to define a ``position'' of the actual focus, so we define it as the average of the two $z$ values at which 90\% of the peak intensity occurs. This operation was repeated for all the measured frequencies and the image position results are summarized in Figure~\ref{fig:Image_Distance}. The overall trends are very similar, again verifying that the metamirror is approximating an ideal focusing mirror quite faithfully. The largest deviation between the ideal focusing mirror and metamirror occurs at 111~GHz where the error is about 15.3\% or 74~mm. Since the images are elongated by this amount or greater, this scale of deviation in image position is not likely to be significant in practice. The observed deviations can again be attributed to the non-ideal $S_{11}$ characteristics of the metasurface. 
\begin{figure}[!h]
  \centering
  \includegraphics[width= 10 cm]{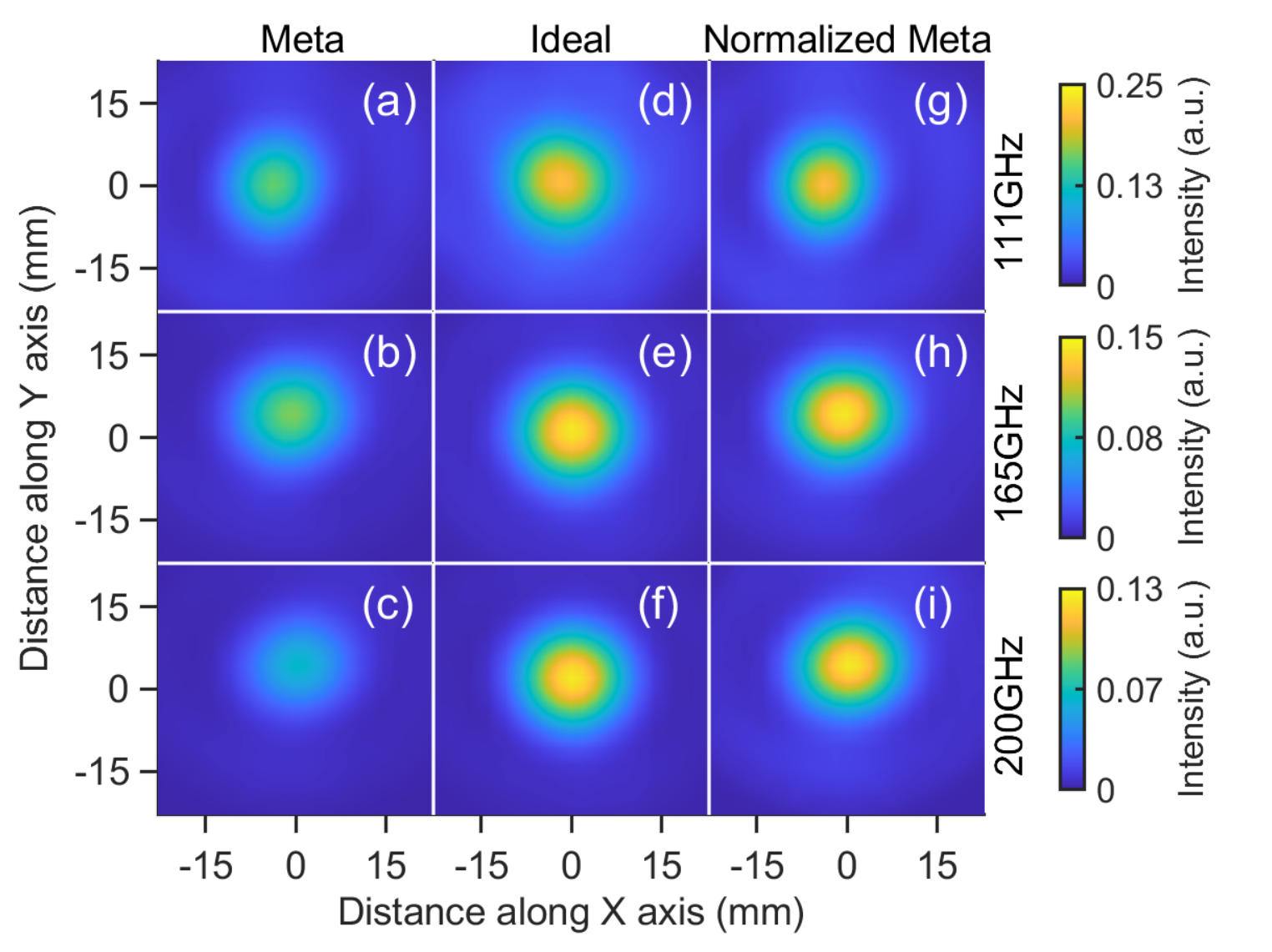}
  \caption{XY plane intensity profiles for metamirror image plane for (a) 111~GHz, (b) 165~GHz, (c) 200~GHz, and XY plane intensity profiles for ideal focusing mirror image plane for 
  (d) 111~GHz, (e) 165~GHz, (f) 200~GHz, and XY plane intensity profiles for metamirror image planes normalized to the peak value of ideal focusing mirror image planes for (g) 111~GHz, (h) 165~GHz, (i) 200~GHz}
  \label{fig:XY_Plane}
\end{figure}

Finally, Figure~\ref{fig:XY_Plane} shows the XY plane intensity profiles for 111, 165, and 200~GHz. Measured XY plane data were interpolated by a factor of 25 times along the Z axis to enable a view of the XY images exactly at their image locations determined above. Individual XY plane data were also interpolated in the $x$ and $y$ directions by a factor of 5.26 times to smooth the appearance of the images. The adoption of this interpolation scheme has been justified by the monotonic and mostly linear change of the measured data (intensity) over the spatial point intervals, which is also justified by simulated intensity profiles. The intensity profiles are not symmetric within the plot windows because the Z axis rail was not exactly aligned with reflected beam axis. Also, the DUT mounting apparatus may have been slightly tilted causing the beams to deflect slightly. Nevertheless, the XY intensity profiles allow us to determine the beam waists and metamirror efficiency. As expected, the intensity profiles for the ideal focusing mirror are brighter than those of the metamirror. To aid comparison, Figs.~\ref{fig:XY_Plane} (g), (h) and (i) show the XY intensity profiles of the metamirror normalized to the corresponding peak value of the XY intensity profiles of the ideal focusing mirror. 

The general size and shape of the metamirror and ideal focusing mirror images are all very similar, again indicating that the metamirror was operating as intended. Compared to the ideal focusing mirror, the metamirror image at 111 GHz reveals a slightly deeper null encompassing the main image lobe. This difference is much less noticeable at higher frequencies. Again, such deviations may be ascribed to the non-ideal $S_{11}$ of the constituent unit cells and possibly the discretization of the metasurface. 

\begin{figure}[!h]
  \centering
  \includegraphics[width= 12 cm]{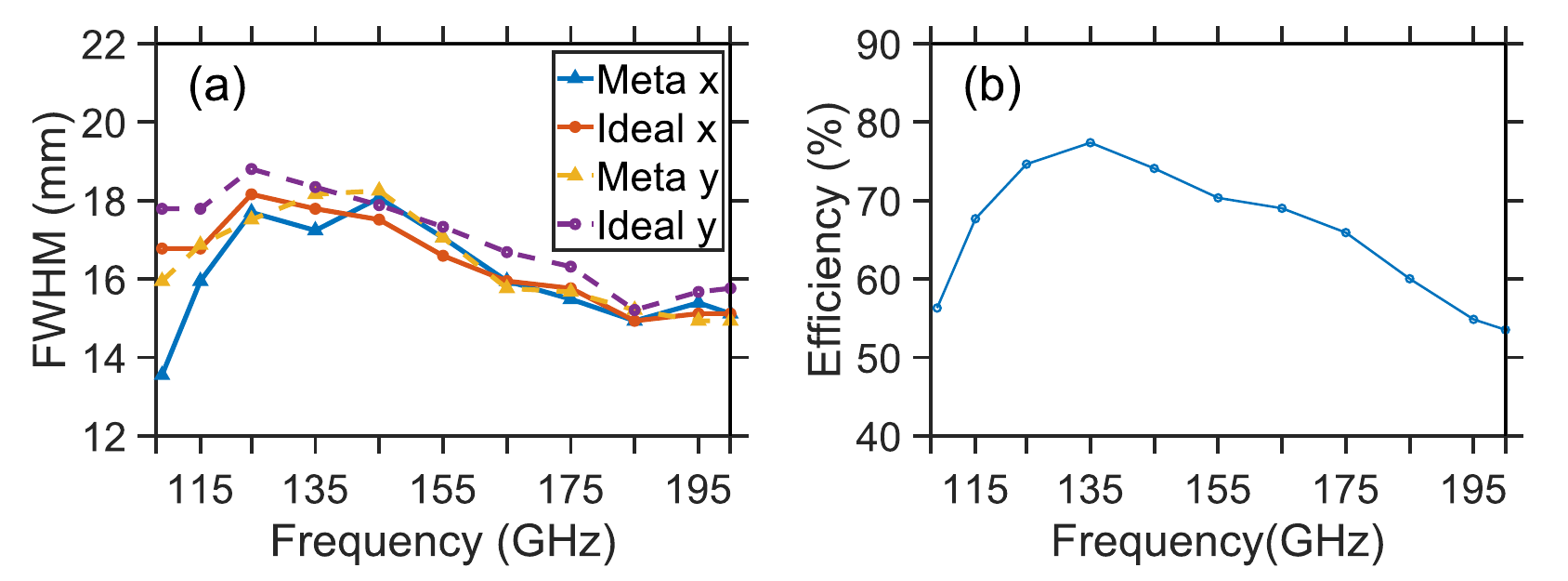}
  \caption{Measured values of (a) FWHM beam waist and (b) energy efficiency.}
  \label{fig:Beam_Eff}
\end{figure}

To provide a quantitative comparison over the entire operating bandwidth, determinations of the full width at half maximum (FWHM) beam waists in both $x$ and $y$ directions were calculated from the measured XY plane intensity profiles. These waist values are presented in Figure~\ref{fig:Beam_Eff}(a). The metamirror beam waists measured in the $x$ direction closely follow those of the ideal focusing mirror over most of the operating bandwidth, with a typical deviation on the order of 5\% (or around 1~mm) or less. The greatest deviation of about 21.2\% or 3.2~mm occurred at 111 GHz. Similar behavior can be observed for the beam waists in the $y$ direction, although in this case the largest deviation (again at 111~GHz) was only 11\% or 1.9~mm. Again, these are mostly insignificant errors in a practical sense, being only a small part of the overall waist size, which in all cases exceeds 13.5~mm. The largest deviations of beam waist values are generally observed at the lower frequencies---this is also likely due to greater non-idealities in $S_{11}$ within that frequency range. 
Comparing the $x$ and $y$ waists of the ideal focusing mirror alone, we see variations of around 6\% or 1~mm in the worst case. This suggests that there is an inherent asymmetry in the experimental setup, which would also manifest in the metamirror images. Indeed, this can be observed in Figure~\ref{fig:XY_Plane}(g) where the nulls near the main lobe in the $x$-direction are more pronounced than those in the $y$-direction. 

The efficiency of the metamirror over the operating band is presented in Figure~\ref{fig:Beam_Eff}(b). To calculate the efficiency, all the intensity values in the XY plane located at the image distance for that frequency were summed. This operation was performed for both the metamirror and ideal focusing mirror, resulting in two single values, $\Sigma I_{\mathrm{Meta_{XY}}}$ and $\Sigma I_{\mathrm{Ideal_{XY}}}$, respectively. The efficiency was then calculated as
 $\Sigma I_{\mathrm{Meta_{XY}}}/\Sigma I_{\mathrm{Ideal_{XY}}}$ \cite{Withayatchumnankul_2019} and plotted. Calculating efficiency in this way, by comparing the metamirror with an ideal focusing mirror of the same target behavior, makes a highly realistic estimation of the metamirror efficiency in real-world applications. From Figure~\ref{fig:Beam_Eff}(b), we observe that the metamirror demonstrates a more than 53.5\% efficiency throughout the intended operating band and peaks as high as 77.4\% at 135 GHz.

\section{Discussion}
\begin{figure}[!h]
  \centering
  \includegraphics[width= 12 cm]{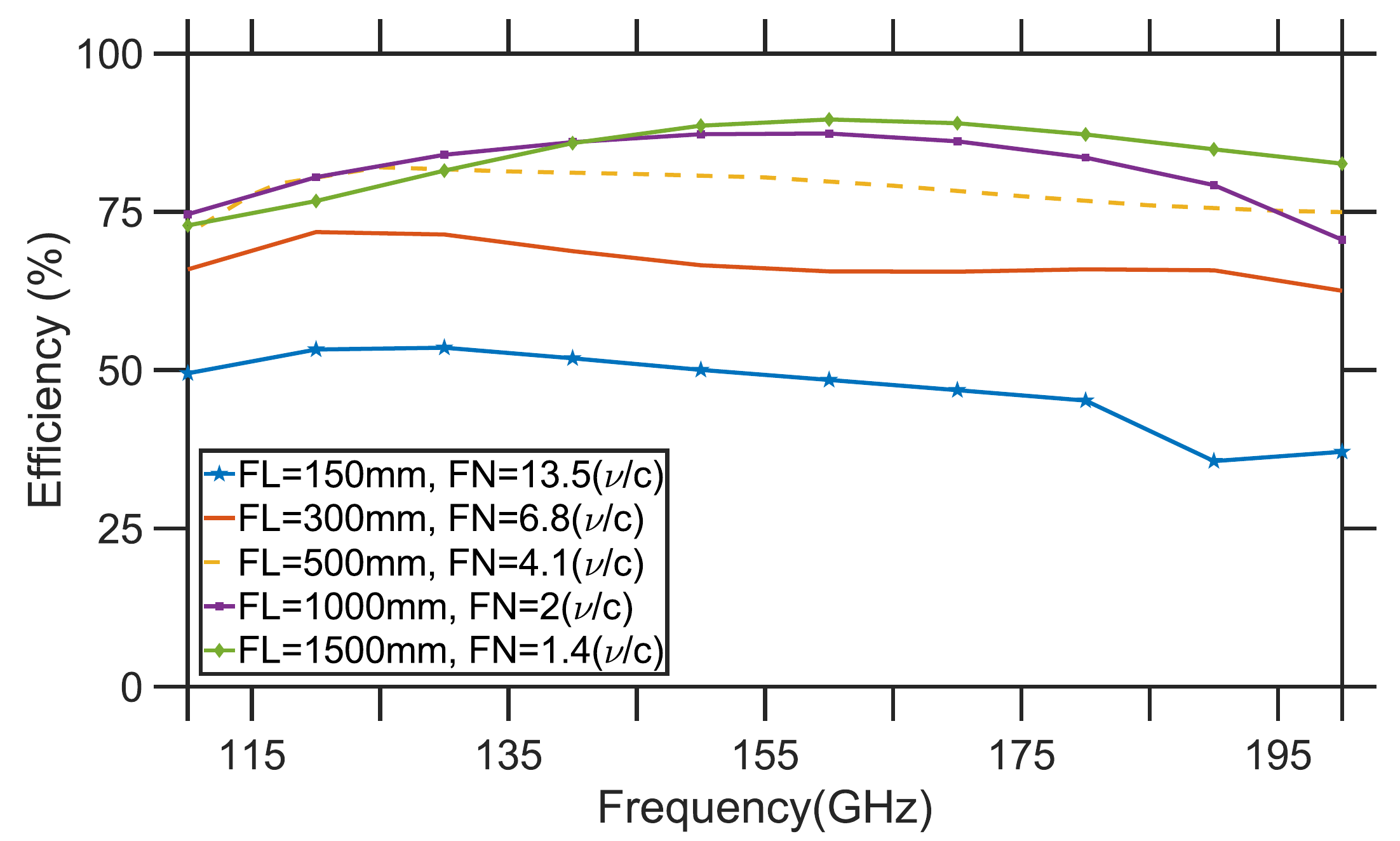}
  \caption{Efficiency vs. frequency curves for simulated metamirrors of different Fresnel numbers. Here, FL and FN mean focal length and Fresnel number respectively. }
  \label{fig:Eff_vs_BW_numerical}
\end{figure}

The high performance of the metamirror compels some important discussion, specifically with regard to the advantages of low Fresnel number optics. Fresnel number describes the number of half period zones in the wavefront amplitude across the mirror aperture radius. Mathematically, if a focusing device having an aperture radius of $\rho$ and focal length of $f$ operates at a wavelength $\lambda$, then Fresnel number (FN) can be expressed as\cite{Parker_1981}
$
FN = {\rho^2}/{(f\lambda)} ,
\label{eqn:2}
$ or equivalently
$FN = {(\rho^2\nu)}/{(cf)}$, where  $c$ is the speed of light and $\nu$ is the operating frequency.
\label{eqn:3}
Despite its simple design, our metamirror performs better than previous designs in terms of combined fractional bandwidth and efficiency, a fact that we attribute to its low Fresnel-number design. To corroborate this claim, we numerically designed four more focusing metamirrors using the same design algorithm and the same pool of unit cells that were used to find our original design. The new mirrors also had the same aperture radius as our original, but the focal lengths were designed at 150~mm, 300~mm, 1000~mm and 1500~mm to investigate a range of Fresnel numbers. The efficiency of these simulated metamirrors was calculated as before, over the 110-200~GHz frequency band, and then plotted in Fig.~\ref{fig:Eff_vs_BW_numerical}. It is observed from Fig.~\ref{fig:Eff_vs_BW_numerical} that the broadband efficiency of the metamirror generally increases when the the Fresnel number is reduced.

Low Fresnel number optics offer several advantages in metamirror design that could explain this behavior. First, they have slower variation of phase across the entire mirror surface. This means it is possible to more coarsely discretize the annular rings that functionalize the metamirror surface, and thus reduce the required number of unit cell designs to achieve a particular performance. Our use of only nine unit cells reveals that finer discretization is not practically necessary in some cases. This can greatly speed the design process since far fewer full-wave simulations are necessary. Second, fewer requried unit cells means it also becomes more probable to find a family of favorable unit cell designs among the finite set of parameterized simulations. This again provides significant relief in the design task, particularly when large bandwidths are involved. Third, low Fresnel number optics are expected to have the most elongated, tubular foci. This means variations in image (or focal) distance are less likely to affect the performance in a significant way, again alleviating design constraints. For example, if an image focus was $>100$~mm in length then it doesn't practically matter if the image location is in error by $\pm25$~mm.   

The matter of image locations warrants further discussion. While the image locations for the original metamirror closely follow those of the ideal focusing mirror over the whole frequency band, they are not what would be expected from a geometric optics analysis. Specifically, our experimental setup predicts geometric image locations at $z=1133$~mm, independent of frequency. Actual image locations range from 450-800~mm on the low and high frequency ends of the bandwidth, respectively. This is not unexpected. 
 Past research shows that optics with low Fresnel numbers demonstrate focal shift whereby the maximum intensity peak appears closer to the focusing optic relative to the geometric focus. This phenomenon has been theoretically, numerically, and experimentally studied in classical optics and is attributed to diffraction effects overwhelming the geometric phase of the focusing wave \cite{Parker_1981, Wolf_1981, Platzer_1983}. Moreover, this phenomenon is general to all focusing systems and has been observed in previous plasmonic- and meta-lens designs \cite{Barnard_2009, Yitting_2011, Shutian_2012, Yuxuan_2018}. Practically, this could be an important design trade-off in future metamirror design considerations.

\section{Conclusion}
 In conclusion, we present a metasurface-based highly efficient broadband metamirror  that operates in the 0.11-0.2~THz regime. We have experimentally demonstrated its performance and also compared it to the performance of an ideal focusing mirror. The performance achieved in this work is, to the best of our knowledge, and with regard to combined efficiency and fractional bandwidth, unmatched in previous experimental demonstrations in the terahertz regime. In terms of beamwidth and image position, our metamirror was shown to behave very closely to an ideal focusing mirror. It can be challenging to achieve such a match over a broad bandwidth because the unit cells are inherently dispersive and a unity reflection magnitude can rarely be obtained. While an ideal focusing mirror obviously performs better, a metamirror can be lighter, thinner and flatter than traditional focusing optics, which makes them more suitable as a foundation for IRSs. Additionally, metasurfaces can be realized using semiconductor materials which may enable their better integration with external optical or electronic systems. While ours is a demonstration of a static metasurface, it reveals important design strategies and mechanisms that can lead to more efficient and \emph{multifunctional or dynamically tunable} metasurfaces. 
 
 Additionally we show that the Fresnel number of meta-focusing devices impacts the achievable fractional bandwidth and energy efficiency. As the Fresnel number grows, so grows the challenge of maximizing these two metrics simultaneously. Taking a more system-oriented viewpoint in design trade-offs, if low Fresnel number optics are a valid option, they may significantly alleviate the design burden. One drawback to note in this work is the frequency dependent focal length of the metamirror, which is exacerbated when using low Fresnel numbers. 

\begin{backmatter}
\bmsection{Funding}
This work was supported in part by the Air Force Office of Scientific Research under Grant FA9550-21-1-0456, the National Science Foundation under Grant ECCS-2238132, and the National Aeronautics and Space Administration under Grant 80NSSC22K0878.

\bmsection{Disclosures}
The authors declare no conflict of interest.

\bmsection{Data availability} Data underlying the results presented in this paper are not publicly available at this time but may be obtained from the authors upon reasonable request.

\bmsection{Supplemental document}
\label{sec:supp}
See Supplement 1 for supporting content. 
\end{backmatter}








\end{document}


\maketitle
\section{Summary of previous research work}
\begin{table}[h]
  \centering
    \caption{Summary of research work previously done on focusing THz metasurfaces. The efficiencies were calculated in various ways in the different works. Fractional bandwidth of $<1$\% has been used where the operating band consists of only a single frequency.The 3$^{\mathrm{rd}}$ column specifies lists the fractional bandwidths (Frac.BW) of the listed devices. The 4$^{\mathrm{th}}$ column specifies whether a polarization conversion (Pol. Conv.) mechanism was employed in the design. The 5$^{\mathrm{th}}$ column lists the estimated Fresnel numbers in various previous works.}
  \begin{tabular}{|c|c|c|c|c|c|}
   \hline
   Device & Efficiency & Frac. BW & Pol. Conv. & Est. Fresnel Num. & Ref. \\
   \hline
   Lens & Not given & 57.14\% & Depn. & 16.7-30 & \cite{Xueqian_2015} \\
   \hline
   Lens & 24\% & $<1$\% & Not depn. & 8.7 &\cite{Ma_2017} \\
   \hline
   Lens & Not given & $<1$\% & Depn. & 300 & \cite{Gaofeng_2018} \\
   \hline
   Lens & $>32\%$, peak 57.5\%& 28.57\% & Depn. & 5-6.7 & \cite{Chunrui_2024}\\
   \hline
   Lens & $>52$\%, peak 65.1\% & 23.22\% & Depn. & 8.3-10.1 & \cite{Chunrui_2024} \\
   \hline
   Lens & $30$\% at 0.65~THz & 30.5\% & Depn. & 7 & \cite{Shixiong_2022} \\
   \hline
   Reflector & $>12$\%, peak 71.9\% & 23.9\% &Not Depn. & 5.8-7.5 & \cite{Withayatchumnankul_2019} \\
   \hline
   Reflector & Not given & 20\% & Depn. & 30.4-37.1 &\cite{Stella_2022} \\
   \hline
   Reflector& Not given & 20\% & Not depn. & 12.6-15.3 & \cite{Headland_2016}  \\
   \hline
   Reflector & $>20$\%, peak 80\% & $28.6$\%& Not depn. & 20.4-27.2 & \cite{Astafev_2015} \\
   \hline
   Reflector & Not given & $9\%$ & Not depn. & 2.1-2.3 & \cite{Dagun_2016} \\
   \hline
   Reflector & Not given & $28.57\%$ & Not depn. & 1.8-2.4 & \cite{Bai_2017} \\
   \hline
   Reflector & $>53.5$\%, peak 77.4\%& 58\%& Not depn. & 1.5-2.7 & This work\\
   \hline
  \end{tabular}

  \label{tab:placeholder}
\end{table}
\section{Focusing reflector working principle}
\begin{figure}[h!]
  \centering
  \includegraphics[width= 11 cm]{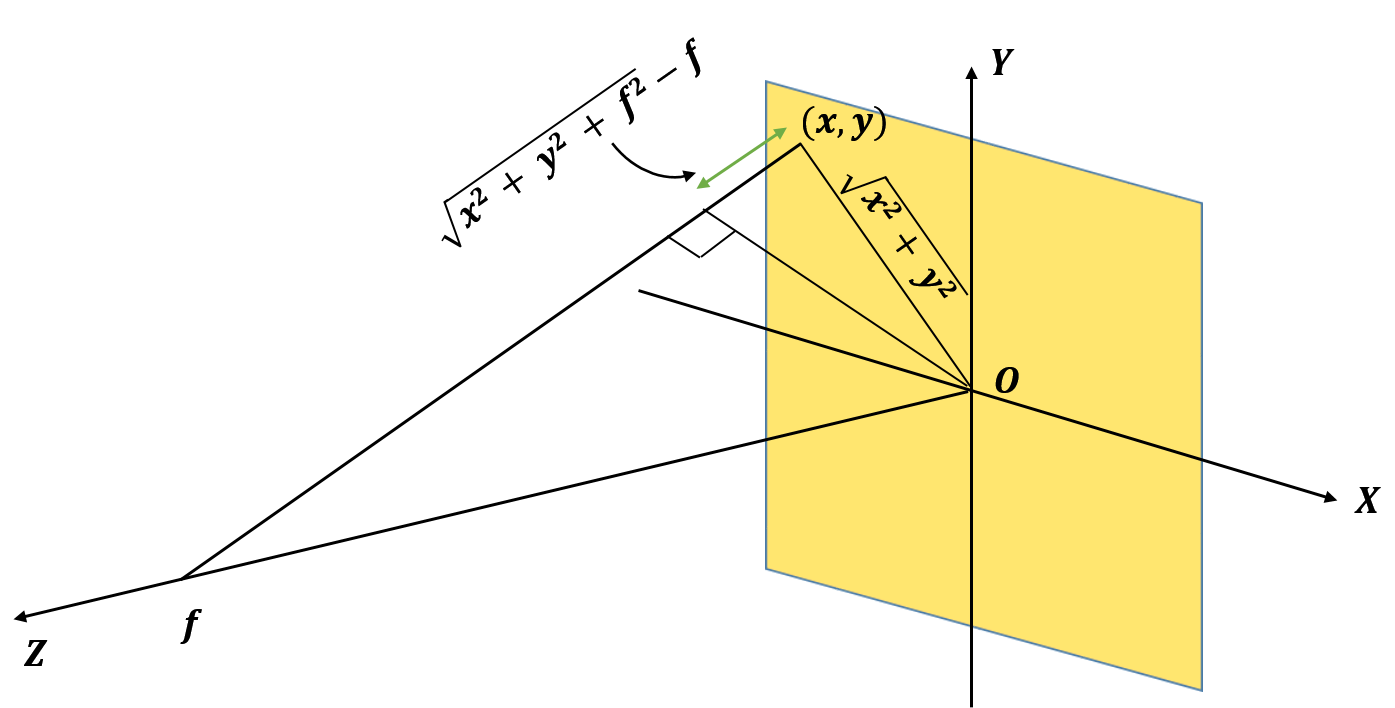}
  \caption{Schematic of required ideal path difference at point (x,y) on the surface of a metamirror.}
  \label{fig:required_ideal_phase}
\end{figure}

A flat focusing reflector or a flat focusing mirror is required to impart a certain phase shift at each point on its surface. The path difference required with the corresponding phase shift at any point can be seen in Figure~\ref{fig:required_ideal_phase}.
\newpage
\section{Unit cell properties}

\begin{figure}[!ht]
\centering
\fbox{\includegraphics[width=1\linewidth]{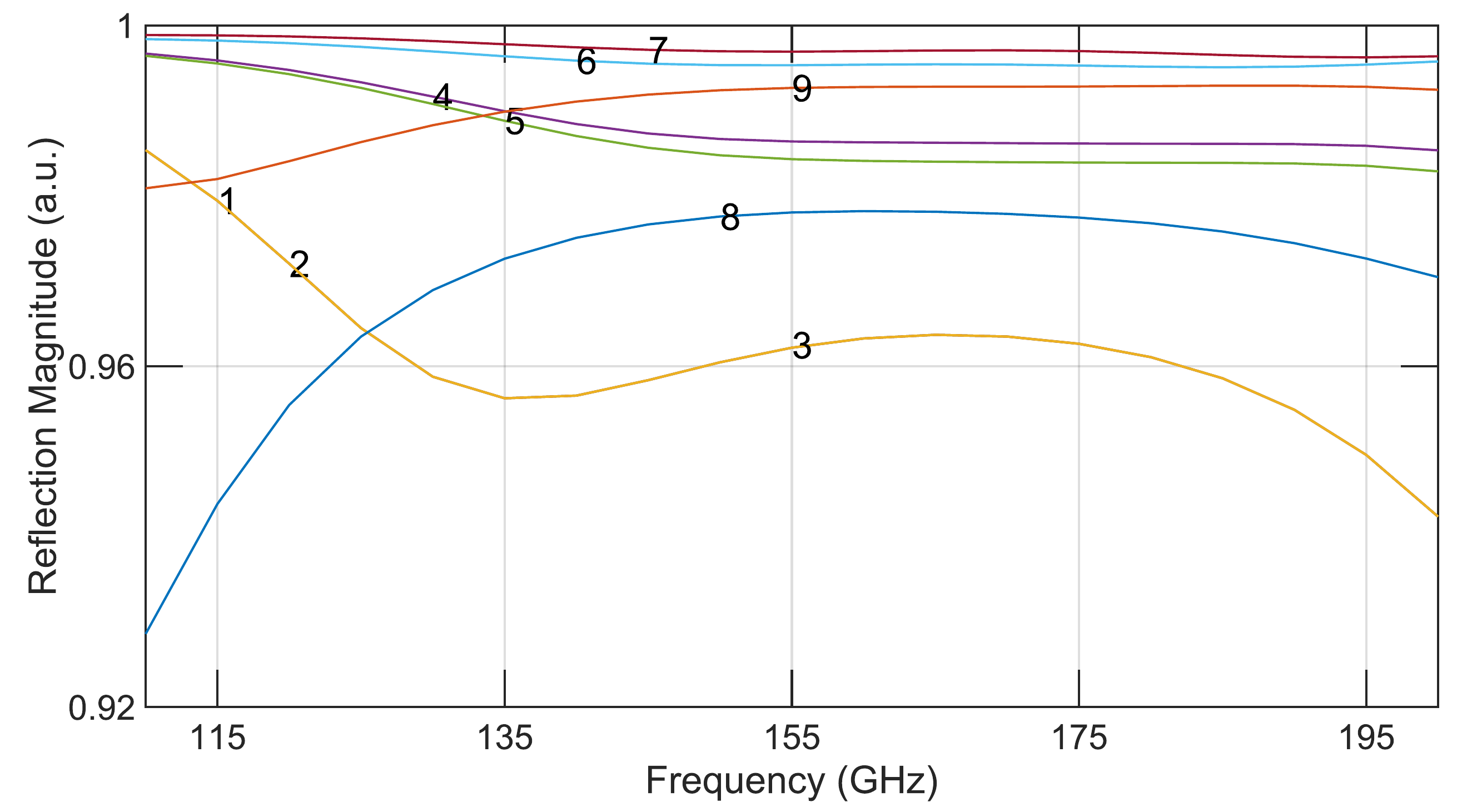}}
\caption{The reflection magnitude of unit cells over the operating frequency band is shown. The numbers 1 to 9 on the curves indicate the sequence of annular regions, beginning at the center and extending towards the circumference of the metamirror. The unit cells 1,2 and 3 are the same and so are their reflection magnitudes.}
\label{fig:mag}
\end{figure}

Figure~\ref{fig:mag} shows the reflection magnitude and Figure~\ref{fig:mag} shows the reflection phase respectively of the unit cells of the metamirror. Table~\ref{tab:dimensions} lists the dimensions of the rectangular gold patches of the unit cells of the metamirror.                                                                                                                                                             
\begin{figure}[!ht]
\centering
\fbox{\includegraphics[width=1\linewidth]{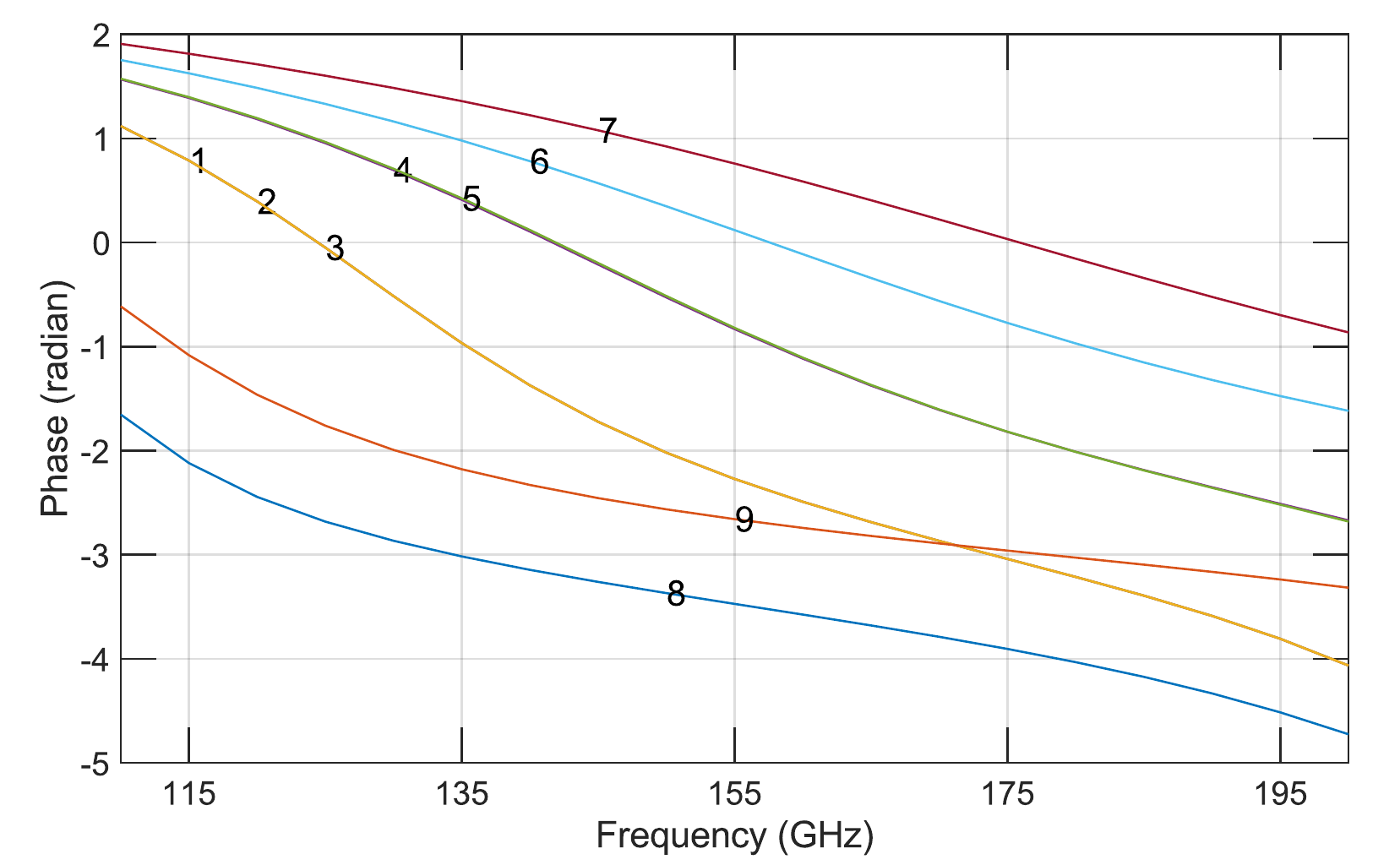}}
\caption{The reflection phase of unit cells over the operating frequency band is illustrated. The numbers 1 to 9 on the curves indicate the sequence of annular regions, beginning at the center and extending towards the circumference of the metamirror.}
\label{fig: phase}
\end{figure}

\begin{table}[!ht]
\centering
\caption{\bf The dimensions of the rectangular patches for the unit cells are listed in this table. The term ‘unit cell no.’ refers to the unit cells that populate the corresponding annular regions in the metamirror. The unit cells 1,2 and 3 are the same and so are their reflection phases. Unit cell 4 and 5 are different but their reflection phase response are very similar. }
\begin{tabular}{|c|c|c|c|c|c|}
\hline
Unit Cell No. & $L1(\mu m)$ & $W1(\mu m)$ & $L2(\mu m)$ & $W2(\mu m)$ & $g(\mu m)$\\
\hline
1 & 465 & 50 & 215 & 120 & 20 \\
\hline
2 & 465 & 50 & 215 & 120 & 20 \\
\hline
3 & 465 & 50 & 215 & 120 & 20 \\
\hline
4 & 90 & 50 & 340 & 120 & 20 \\
\hline
5 & 340 & 50 & 340 & 50 & 20\\
\hline
6 & 90 & 50 & 215 & 260 & 20\\
\hline
7 & 90 & 50 & 90 & 50 & 20\\
\hline
8 & 90 & 50 & 590 & 50 & 20 \\
\hline
9 & 465 & 260 & 90 & 260 & 20\\
\hline
\end{tabular}
  \label{tab:dimensions}
\end{table}